\documentclass[graybox]{svmult}

\usepackage[utf8]{inputenc}
\usepackage[english]{babel}

\usepackage{lipsum}
\usepackage[T1]{fontenc}
\usepackage{mathtools}
\usepackage{standalone}
\usepackage{shellesc}
\usepackage{setspace}
\usepackage{wrapfig}
\usepackage{textcomp}
\usepackage{amssymb}
\usepackage{amsmath}
\usepackage{scalerel,stackengine}

\usepackage{ifthen}
\usepackage{bm}
\usepackage{mathptmx}
\usepackage{helvet}
\usepackage{courier}
\usepackage{type1cm}

\usepackage[sort]{natbib}
\usepackage{makeidx}
\usepackage{multirow}
\usepackage{graphicx}

\usepackage{multicol}
\usepackage[bottom]{footmisc}
\usepackage{siunitx}
\usepackage{xspace}

\usepackage{hyperref}
\usepackage[capitalise]{cleveref}

\usepackage{xfrac}
\usepackage{tikz,pgfplots}
\usepackage{tikzscale}
\usepackage{subcaption}
\usepackage{booktabs}

\usepackage{makecell}
\usepackage[ruled,vlined]{algorithm2e}
\crefname{algocf}{algorithm}{algorithms}
\Crefname{algocf}{Algorithm}{Algorithms}

\DeclareSIUnit\flops{FLOPs}

\usetikzlibrary{calc}
\usetikzlibrary{plotmarks,patterns}
\usepgfplotslibrary{fillbetween,groupplots}
\usepgfplotslibrary{colormaps}
\usepgfplotslibrary{colorbrewer}
\usetikzlibrary{colorbrewer}
\usetikzlibrary{shapes.geometric}
\usetikzlibrary{arrows}
\pgfplotsset{
	compat=newest,
  cycle list/Dark2,
}
\usetikzlibrary{external}
\usepgfplotslibrary{external}
\newboolean{dopdflatex}

\setboolean{dopdflatex}{true}

\newcommand\ifpdflatex[2]{\ifthenelse{\boolean{dopdflatex}}{#1}{#2}}

\newcommand{\code}[1]{\texttt{#1}}

\newcommand{\flexi}{FLEXI\xspace}

\newcommand{\galaexi}{GA\-L{\AE}\-XI\xspace}

\newcommand{\numThreePlaces}[1]{\num[round-mode=places,round-precision=3]{#1}}

\mathchardef\mhyphen="2D

\definecolor{mycolor1}{HTML}{1b9e77}
\definecolor{mycolor2}{HTML}{d95f02}
\definecolor{mycolor3}{HTML}{7570b3}
\definecolor{mycolor4}{HTML}{e7298a}
\definecolor{mycolor5}{HTML}{66a61e}
\definecolor{mycolor5}{HTML}{e6ab02}
\definecolor{mycolor5}{HTML}{a6761d}
\definecolor{mycolor6}{HTML}{666666}

\makeindex

\begin{document}

\title*{APU-Accelerated Large Eddy Simulation with the Discontinuous Galerkin Solver \galaexi}
\titlerunning{APU-Accelerated Large Eddy Simulation with \galaexi}

\author{
  \mbox{Spencer Starr}$^\dag$,
  \mbox{Anna Schwarz}$^\dag$,
  \mbox{Justin Du Plessis}$^\dag$,
  \mbox{Andreas Wanninger}$^\dag$,
  \mbox{Johanna Hintz}$^\dag$,
  \mbox{Rohan Kaushik}$^\dag$,
  \mbox{Patrick Kopper}$^\dag$,
  and \mbox{Andrea Beck}$^\dag$
}
\authorrunning{
  Starr,
  Schwarz,
  Du Plessis,
  Wanninger,
  Hintz,
  Kaushik,
  Kopper,
  and Beck
}

\institute{$^\dag$ Institute of Aerodynamics and Gas Dynamics, University of Stuttgart, \\
\\ \email{
  starr/
  schwarz/
  du-plessis/
  wanninger/
  hintz/
  kaushik/
  kopper/
  beck@iag.uni-stuttgart.de
}}

\maketitle
\abstract{
  The exascale computing era, driven by heterogeneous GPU architectures, requires a fundamental redesign of traditional CFD solvers to fully leverage those heterogeneous systems. The discontinuous Galerkin spectral element method (DGSEM) provides an ideal foundation for this transition due to its high-order accuracy and local computational stencil. This work presents recent advances in the development and application of the architecture-agnostic DGSEM framework \galaexi by linking hardware optimization, software implementation, and physical validation. The performance of \galaexi on the AMD MI300A Accelerated Processing Units (APUs) featured on the Hunter supercomputer is analyzed. Specifically, evaluations of the strong and weak scaling performance and the impact of the compute partitioning modes available on the AMD MI300As are performed. Second, the strategy used to integrate the algorithms necessary for wall-modeled large eddy simulations into the GPU-accelerated framework is outlined. Validation of those algorithms is presented in the form of a plane turbulent channel testcase. Finally, the solver is applied to a demanding flow problem in the form of a wall-resolved large eddy simulation of a transonic compressor cascade. The results from this investigation demonstrate the capabilities of \galaexi to accurately capture complex shock-wave/turbulent boundary-layer interactions.
}

\section{Introduction}
\label{sec:introduction}

Over the last several decades, the steady increase of computational resources has transformed the computational sciences, allowing researchers to tackle increasingly complex research questions through larger and more impactful simulations. This new generation of supercomputers offers unprecedented computing power that promises to accelerate both fundamental scientific research and industrial innovation. Unlike previous generations, the performance of exascale systems is primarily driven by heterogeneous architectures centered on GPU-based accelerators. While these systems offer massive throughput, their stream-based nature renders branching operations and host-to-device data transfers relatively expensive.

Consequently, the design and optimization of computational fluid dynamics (CFD) solvers must inherently adapt to these hardware
shifts to realize potential performance gains. High-order numerical methods, particularly the discontinuous Galerkin spectral
element method, provide an ideal foundation for this transition due to their high-order accuracy, local computational stencils, and
efficient memory utilization. The DGSEM constitutes the basis for the CPU-based CFD framework \flexi~\cite{krais2020flexi}, as well
as its accelerated version \galaexi~\cite{kurz2024galaexi,starr2026glxi}. Both of these are currently under active development and have been applied to a variety of complex applications, including aeroacoustics, particle-flow
interactions, and turbulent flows~\cite{beck2018application,kempf2021development,blind2023towards,Kopper2023,Schwarz2025,Keim2026}.

This report addresses the recent advances in the exascale transition of \galaexi by bridging hardware-specific optimization, efficient software implementation, and high-fidelity physical validation. To this end, the report is structured across three critical pillars. In \cref{sec:performance}, the GPU-acceleration strategy of \galaexi is briefly outlined, alongside the parallel performance and a comprehensive investigation into the effect of partitioning modes on AMD MI300A Accelerated Processing Units (APUs). Building on these hardware insights, the GPU-acceleration of the algorithms necessary for wall-modelled large eddy simulations (WMLES) in \galaexi is covered in~\cref{sec:wm}. Parallel to this computational advancement, the framework is applied to a physically demanding wall-resolved large-eddy simulation (WRLES) of a transonic compressor cascade, investigating the shock-wave/turbulent boundary-layer interaction (STBLI) at the suction side of the blade. This interaction serves as a rigorous test case, requiring the solver to capture both the fine-scale turbulence of the boundary layer and the sharp discontinuities of the shock system. \cref{sec:summary} summarizes the findings of this work.

\section{Performance}
\label{sec:performance}

Modern heterogeneous HPC architectures rely heavily on accelerators to tackle large-scale problems. Leveraging these systems often requires the redesign of existing software and special implementations of GPU-enabled programming models. The following section discusses the strategy used in \galaexi to support GPU-accelerated computations on accelerator-based systems, with specific focus the AMD MI300A APU architecture featured in the Hunter supercomputer at HLRS. Discussions then follow on the performance of \galaexi on that hardware, including strong and weak scaling and a comparison of the different compute partitioning modes available on the AMD MI300A.

\subsection{Strategy for GPU-Acceleration}
\label{subsec:porting}

For GPU-accelerated simulations on the AMD MI300A APUs, compute kernels in \galaexi are implemented in HIP C++~\cite{hiplatest}. The Fortran core of \galaexi is interfaced to these compute kernels using Fortran-C interfaces as defined in the Fortran standard~\cite{reid2007fortran}. All operations that mutate the memory of the GPU (i.e. allocate or copy) are issued from the HIP C++ API. Fortran code calls a wrapper which passes to a C++ function that then uses the HIP API (e.g. \code{hipMalloc}). A pointer to allocated GPU memory is stored in a Fortran variable, where it is passed to HIP kernels later to perform operations on the data. CPU-allocated Fortran arrays are mapped to flattened, one-dimensional arrays in GPU memory and data retains Fortran column major ordering in HIP kernels. Indexing functions are used in kernel code to find offsets in the flattened arrays to the data assigned to each GPU thread.

The algorithms in \galaexi are parallelized in HIP kernels by assigning a single GPU thread to operate on a single degree of freedom (DOF). Beyond the kernel level, \galaexi employs \emph{device streams}~\cite{hiplatest} to allow multiple, independent operations in the DGSEM operator execute on a single GPU simultaneously. \galaexi uses three device streams with three levels of priority. The highest priority is given to kernels that perform computations on element faces at the boundary between MPI processes. The lowest priority is for volume operations within elements, which are independent of other neighboring elements. To control data dependencies between kernels executing on different streams, \galaexi uses \emph{device events} for synchronization~\cite{hiplatest}. GPU-aware MPI is leveraged on distributed systems to scale \galaexi to multiple GPUs. The computational domain is divided into non-overlapping subdomains using a space-filling curve and each subdomain is assigned to a single MPI process, which corresponds to a CPU-core-GPU pair.
The approach for GPU-acceleration used in \galaexi is described in further detail in \cite{starr2026glxi}.

\subsection{Scaling Performance}
\label{subsec:scaling}

\begin{figure}
  \begin{center}
    \includegraphics[width=0.8\textwidth]{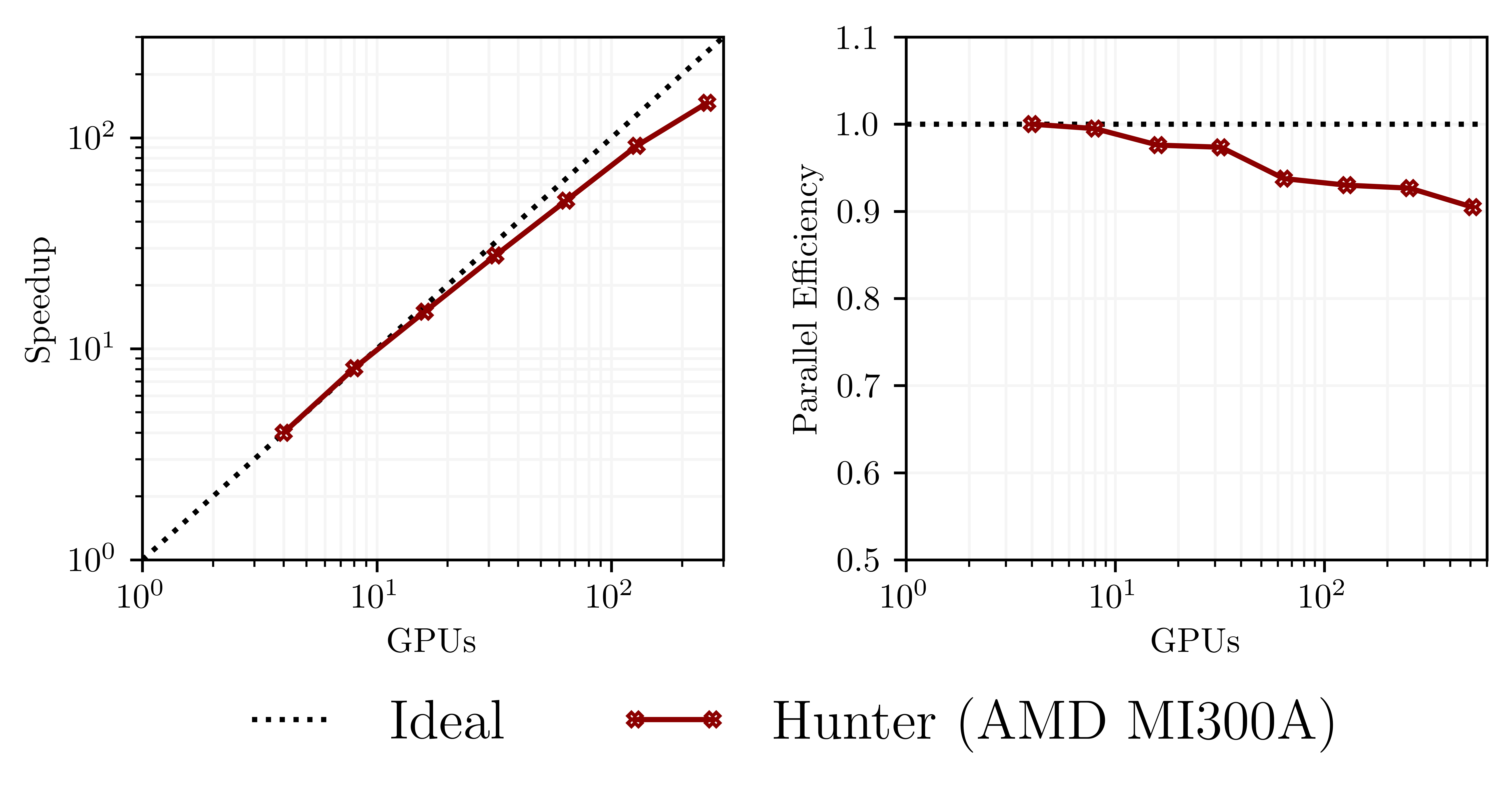}
  \end{center}
  \caption{Strong (left) and weak (right) scaling of \galaexi on Hunter. The ideal speedup and parallel efficiency are indicated by the black dotted lines.}
  \label{fig:scaling}
\end{figure}

To demonstrate the solver's scaling capabilities on Hunter, a verification test case is conducted using a uniform solution initialized on a periodic 3D Cartesian mesh. As \galaexi relies on explicit time-stepping, the performance is entirely independent of the initialized solution. For the scaling tests, a split-form DG scheme with a polynomial degree of $N=9$ evaluated at Legendre--Gauss--Lobatto nodes was employed. Each computation was advanced in time for 10 explicit Runge--Kutta (RK) time steps. The strong and weak scaling performance of \galaexi on Hunter are illustrated in \cref{fig:scaling}. The strong scaling experiments were performed up to 64 nodes (256 MI300A APUs) with a computational domain featuring 524 million DOFs. \galaexi showcases excellent scalability on Hunter, achieving near ideal strong scaling. The weak scaling was measured up to 128 nodes. For the weak scaling runs, the computational domain is incrementally increased to maintain a constant load of 4.096 million DOFs per APU. In the largest weak scaling case, a mesh of 2.097 billion DOFs was run on 512 APUs with a parallel efficiency of 90.4\%. Strong scaling speedup and weak scaling parallel efficiency are both calculated with respect to the single node wall times.

\subsection{Effect of AMD MI300A APU Partitioning Modes}
\label{subsec:partitioning}

The AMD MI300A APUs offer the ability for users to configure the compute resources of the chip to meet application-specific needs. The compute units (CUs) of a single AMD MI300A APUs are built on the chip as six sub-chips, called Accelerated Complex Dies (XCDs)~\cite{amddriverlatest}. It is possible for the six XCDs to be partitioned in three different modes. Single Partition Accelerator (SPX) mode is the manufacturer default which treats all six XCDs as a single monolithic GPU~\cite{amddriverlatest}. It is presented as the preferred mode for applications that benefit from the large throughput and memory bandwidth available from a full MI300A. The Triple Partition Accelerator (TPX) and Core Partition Accelerator (CPX) modes then subdivide the XCDs, creating three logical GPUs in TPX mode and six in CPX mode. CPX mode can offer better performance when applications run many smaller tasks in parallel and TPX mode provides a balanced, middle option.

All scaling performance results presented in \cref{subsec:scaling} were collected using SPX mode. However, in typical applications, \galaexi demonstrates neither of the features (i.e. large memory and compute requirements) to fully benefit from SPX mode. The explicit DGSEM implemented in \galaexi requires roughly 1.2 GB of RAM per 1 million DOFs, resulting in a typical production-scale run on AMD MI300As (i.e. roughly 4 million DOFs/APU) using 3.75\% of the available RAM. For simulations on Hunter at the same 4 million DOFs/APU, \galaexi produces a maximum compute throughput ($\text{R}_{max}$) that is 2\% of the 61.3 TFLOP peak FP64 throughput ($\text{R}_{peak}$) of the MI300A~\cite{mi300a}. The main limiting factor in \galaexi to pushing utilization higher is the highly memory bound nature of the DGSEM algorithms when implemented for GPUs.

\begin{table*}[h]
  \centering
  \caption{
    Comparison of the performance of \galaexi with the three available compute partitioning modes (SPX, TPX and CPX) for the AMD MI300A APUs.
  }
  \label{tab:modes}
  \begin{tabular}{l||c|c|c|c|c|c}
    \toprule
    \multirow{2}{*}{\thead{DOFs/APU}} & \multicolumn{3}{c|}{\thead{PID [s]}} & \multicolumn{3}{c}{\thead{Time-to-solution [s]}} \\
                                        & \thead{SPX} &  \thead{TPX} & \thead{CPX} & \thead{SPX} &  \thead{TPX} & \thead{CPX}\\
    \hline
    \num{7.02e5}   & \numThreePlaces{4.260270E-09} & \numThreePlaces{3.467800E-09} & \numThreePlaces{2.984217E-09} & \num{4.12}  & \num{2.76}  & \num{2.66} \\
    \num{1.40e6}  & \numThreePlaces{3.207810E-09} & \numThreePlaces{3.194587E-09} & \numThreePlaces{2.812900E-09} & \num{5.01}  & \num{2.68}  & \num{1.88} \\
    \num{2.80e6}  & \numThreePlaces{2.945070E-09} & \numThreePlaces{2.937840E-09} & \numThreePlaces{2.731667E-09} & \num{8.74}  & \num{4.01}  & \num{3.11} \\
    \num{5.62e6}  & \numThreePlaces{2.800100E-09} & \numThreePlaces{2.774193E-09} & \numThreePlaces{2.720517E-09} & \num{16.45} & \num{7.55}  & \num{5.44} \\
    \num{1.12e7} & \numThreePlaces{2.635230E-09} & \numThreePlaces{2.632037E-09} & \numThreePlaces{2.539633E-09} & \num{32.3}  & \num{14.68} & \num{10.31} \\
    \num{2.25e7} & \numThreePlaces{2.564350E-09} & \numThreePlaces{2.550310E-09} & \numThreePlaces{2.456583E-09} & \num{63.89} & \num{27.59} & \num{20.42} \\
    \bottomrule
  \end{tabular}
\end{table*}

From the above factors, the TPX and CPX modes, which are oriented toward lightweight applications, are predicted to offer some amount of performance benefit to \galaexi. To test the effect of the partitioning mode on the performance of \galaexi, eight nodes (32 MI300A APUs) on Hunter were used to compute a series of simulations with progressively larger meshes. A freestream flow in a Cartesian box was chosen as the testcase and the polynomial order of the DGSEM was set to $N=7$. The results of these experiments are presented in \cref{tab:modes} with two performance metrics, the performance index (PID) per APU and the time-to-solution. PID is calculated as
\begin{equation}
	\mathrm{PID} = \frac{\mathrm{wall}\mhyphen\mathrm{clock}\mhyphen\mathrm{time} \cdot \mathrm{\#APUs}}{\mathrm{\#DOF} \cdot \mathrm{\#time\;steps} \cdot \mathrm{\#RK}\mhyphen\mathrm{stages}} ,
 \label{eq:pid}
\end{equation}
noting that, regardless of the partitioning mode used, the total number of full \emph{APUs} is used in \cref{eq:pid}. Time-to-solution is the total wall time of a computation, including all wall time required for initialization, data analysis and file output.

CPX mode produces an average reduction in the PID of 10.03\% over SPX mode across the range of mesh sizes, while TPX mode provides negligible benefit. As \galaexi is a memory-bound application, the source of the improvement with CPX mode comes from better memory locality and cache reuse in algorithms that require large numbers of non-contiguous memory accesses. Between SPX mode and TPX mode, speedups in time-to-solution of 1.86x to 2.31x speedups are seen for all cases featuring a computational load of 1 million DOFs/APU or larger. The benefits are greater for CPX mode, which produced speedups between 2.66x and 3.13x in time-to-solution over SPX mode in the same range.

\section{GPU-Accelerated Wall-Modeled LES}
\label{sec:wm}

Despite ever-growing high performance computing resources, turbulent wall-boun-\\ded flows at high Reynolds numbers imply prohibitively high computational costs when solved with WRLES. The restrictive resolution constraints of WRLES are driven by the near-wall problem, which dictates that the size of energy-carrying eddies in the inner boundary layer scales down with increasing Reynolds numbers. Consequently, Deardorff~\cite{Deardorff1970} proposed to model the inner part of the boundary layer, while still resolving the rest of the domain to circumvent the near-wall problem \cite{Kawai2012}. To this day, the associated WMLES approach provides the optimal trade-off between accuracy and computational cost for high Reynolds number wall-bounded flows.

The following section outlines the algorithms necessary for integrating WMLES into \galaexi. Prior to discussing the implementation, a brief overview of wall modeling approaches is provided to contextualize the selected wall model. While the underlying theory is provided for completeness, the primary focus of this work is on the GPU-accelerated implementation within \galaexi. The GPU-acceleration strategy is subsequently validated using a turbulent channel flow benchmark.

\subsection{Wall Modeling}
A plethora of WMLES approaches have been proposed in recent decades, whereof the focus in this study is placed on wall-stress models, following the taxonomy introduced by Larsson et al.~\cite{Larsson2016}.
Physics-based wall stress models formally solve the large-eddy simulation (LES) all the way down to the wall and rely on either the ensemble-/Favre-averaged or low-pass filtered Navier--Stokes--Fourier equations as governing equations to model the effect of unresolved energetic scales in the inner layer \cite{Larsson2016}.
Two of the most prominent categories of wall-stress models are algebraic models, e.g. based on Spalding’s law and ODE-based models.
Algebraic wall-stress models rely on a functional relationship between the velocity and wall-normal distance in viscous units. Thus,
they bypass the need to solve the filtered Navier--Stokes equations (PDE-based models) or the thin boundary layer equations
(ODE-based models), which assume the effects of convective transport and pressure gradients cancel out under the equilibrium
assumption. Consequently, algebraic models offer a significantly reduced computational cost compared to their PDE- and ODE-based
counterparts. However, their accuracy is typically constrained by the underlying assumptions of both flow equilibrium and
incompressibility, though more recent variants have been developed to address these shortcomings such as by considering non-equilibrium effects~\cite{Yang2015}.
While for many engineering applications the equilibrium assumption is applicable, such as the turbulent channel flow test case
considered in this study, the incompressibility assumption can be relaxed through the use of a Van-Driest velocity
transformation~\cite{Van_Driest:1951,White:2006}, which allows to map the incompressible to the compressible boundary layer.

The coupling between the LES domain and the wall model is executed through the following mechanisms. For a given interface point located at a distance $h_{\mathrm{wm}}$ from the wall, the wall model samples the instantaneous LES data. Based on these local flow characteristics, the wall model computes a wall shear stress \(\tau_w\), and, for the case of compressible flows with an isothermal wall boundary condition, additionally the wall heat flux \(q_w\). These values are subsequently imposed on the viscous fluxes at the wall. To facilitate this, the no-slip boundary condition employed in WRLES is relaxed to a slip condition, thereby leaving the tangential velocity components unconstrained. This is due to the fact that the respective wall model enforces the no-slip condition either as a boundary condition, or, in the case of algebraic wall-stress models, by construction. While alternative enforcement strategies have emerged recently, the present work is constrained to the classical slip boundary condition. The viscous fluxes are governed by the computed wall shear stress and, optionally, the wall heat flux.

\subsection{GPU-Acceleration of Wall-Models for Compressible Flows}
Implementing the features necessary for WMLES in to \galaexi is essential, as the use of GPU-accelerated architectures allows access to large-scale WMLES production cases that were previously prohibitive in size. To this end, a physics-based, algebraic, equilibrium wall-stress model based on Spalding's law of the wall is utilized, which reads as
\begin{equation}
y^+
=
u^+
+
0.1108
\left[
e^{0.4u^+}
- 1
- 0.4u^+
- \frac{\left(0.4u^+\right)^2}{2!}
- \frac{\left(0.4u^+\right)^3}{3!}
- \frac{\left(0.4u^+\right)^4}{4!}
\right]
\end{equation}
and results in a unified expression for the streamwise velocity \(u^+\) in the viscous sublayer, buffer layer, and the log layer in
terms of the wall-normal distance \(y^+\). Note that both streamwise velocity \(u^+\) and wall-normal distance \(y^+\) are given in viscous wall units which allows solving this nonlinear equation for the wall shear stress \(\tau_{w}\). The wall model can be extended to compressible flows by computing an equivalent incompressible velocity. The equivalent velocity is based on the instantaneous, compressible LES velocity at the wall model interface through the aforementioned Van--Driest velocity transformation. See~\cite{blind2023towards} for further details.

Implementing the equilibrium-based compressible algebraic wall-stress model relies on several underlying utilities. The interface utility provides primitive solution variables at the matching height $h_{wm}$ and the weak enforcement of wall shear stress within the viscous fluxes. Coverage of these foundational routines is omitted here and the focus is laid on the acceleration strategy for the core wall-modeling routine itself. Notably, the following formulation implies that the wall model is strictly applicable to compressible flows featuring an adiabatic wall boundary condition. An overview of the original CPU-based approach is provided in Algorithm~\ref{alg:CPU}, with its GPU counterpart presented in Algorithm~\ref{alg:GPU}.

As detailed in Algorithm~\ref{alg:CPU}, the legacy CPU framework processes the wall model using a nested loop hierarchy. The solver first initiates a top-level loop over all boundary faces designated for wall modeling, specifically, those sides where the fluid volume is intercepted by the wall-normal interface height $h_{wm}$. For every wall-modeled face, a nested loop iterates over all local DOFs to retrieve the primitive variables in local wall-normal coordinates. For the algebraic model of interest, only the wall-parallel velocity components are extracted to compute the streamwise velocity magnitude.
This velocity magnitude serves as the direct input for the core wall model logic, which effectively acts as a black box to output
the wall-shear stress magnitude, which is then projected along both tangential directions and weakly enforced inside the viscous fluxes at the wall.
Operationally, the model execution follows a multi-step numerical sequence at each DOF, applying the Van Driest transform and executing the nonlinear root-finding solver.

Within \galaexi the core convention dictates that each individual DOF per wall-modeled face is mapped to a single GPU thread. Rather than sequentially stepping through all boundaries and their associated DOFs, the GPU implementation processes them concurrently, cf. Algorithm~\ref{alg:GPU}. Consequently, the legacy nested loop structure is entirely replaced by a unified kernel function call implemented in HIP C++. Inside the kernel itself, the overarching philosophy of \galaexi is preserved: the original, validated CPU-based Fortran logic is modified only where strictly necessary for hardware acceleration. As a result, the fundamental steps of the routine—recovering the primitive variables, applying the Van Driest transform, and executing the nonlinear root-finding solver—remain numerically identical to their CPU counterparts, ensuring strict consistency across architectures.

\begin{algorithm}[t]
\caption{CPU-based wall-model implementation.\label{alg:CPU}}
\DontPrintSemicolon
\SetKwFunction{WallModel}{WallModel}
\SetKwFunction{VD}{VanDriestTransform}
\SetKwFunction{SolveTau}{SolveWallStress}

\ForEach{wall-modeled face $f$}{
  \ForEach{degree of freedom $(i,j)$ on face $f$}{
    Compute $\tau_w \leftarrow \SolveTau(\tilde{u}_{\parallel}, h_{\mathrm{wm}})$\;
  }
}
Weakly enforce component-wise wall shear stress $\tau_w$ in viscous fluxes\;
\end{algorithm}

\begin{algorithm}[t]
\caption{GPU-based wall-model implementation. \label{alg:GPU}}
\DontPrintSemicolon
\SetKwFunction{Kernel}{WallModelKernel}
\SetKwFunction{VD}{VanDriestTransform}
\SetKwFunction{SolveTau}{SolveWallStress}

Launch \Kernel with one thread per face degree of freedom\;

\ForEach{GPU thread assigned to $(f,i,j)$}{
  Compute flattened-array offsets\;
  Compute streamwise velocity magnitude $u_{\parallel}$ at $h_{\mathrm{wm}}$\;
  Compute $\tau_w \leftarrow \SolveTau(\tilde{u}_{\parallel}, h_{\mathrm{wm}})$\;
}
Weakly enforce component-wise wall shear stress $\tau_w$ in viscous fluxes\;
\end{algorithm}

\subsection{Validation of the Ported Wall Model}
The GPU-accelerated implementation of Spalding's algebraic wall model is validated using a plane turbulent channel flow at Mach and Reynolds numbers of $\mathrm{Ma}=0.1$ and $\mathrm{Re}_{\tau}=180$, which serves as a common benchmark case for validating equilibrium-based wall models.
The computational domain $\Omega \in [0,2\pi] \times [0,2\delta] \times [0,\pi]$ with channel half-height $\delta=1$, is discretized using \((5,30,3)\) elements in the respective directions, where the mesh spacing in the wall-normal ($y$) direction decreases successively towards the wall.
A polynomial degree of \(N=5\) is chosen, which results in a total number of \num{162000} DOFs.
Periodic boundary conditions are imposed in the streamwise and spanwise directions. The aforementioned slip boundary condition,
including enforcement of the wall shear stress in the viscous fluxes computed via the wall model, are prescribed on the bottom and upper wall.
Initially, an analytical mean turbulent velocity profile of constant, unit density is assigned. Sine perturbations are superimposed in all three velocity components in order to imitate turbulent structures. Additionally, a constant pressure gradient \( \frac{\partial p}{\partial x}=-1 \) is used as a volume source term to avoid slow-down of the flow.

\begin{figure}[htb]
    \centering
    \begin{subfigure}[t]{0.75\columnwidth}
        \centering
        \includegraphics[width=\linewidth]{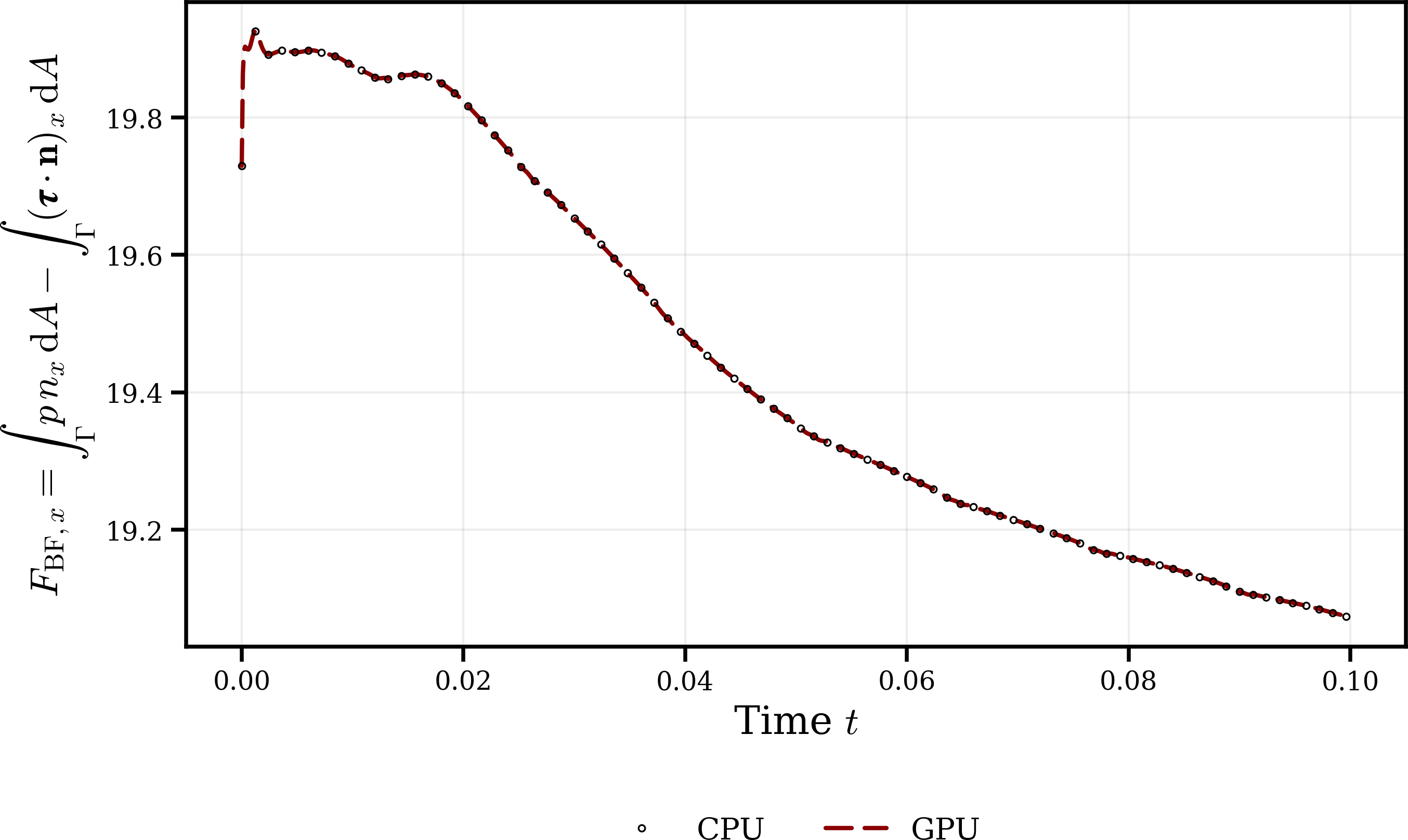}
        \label{fig:wm_force}
    \end{subfigure}
    \hfill
    \begin{subfigure}[t]{0.75\columnwidth}
        \centering
        \includegraphics[width=\linewidth]{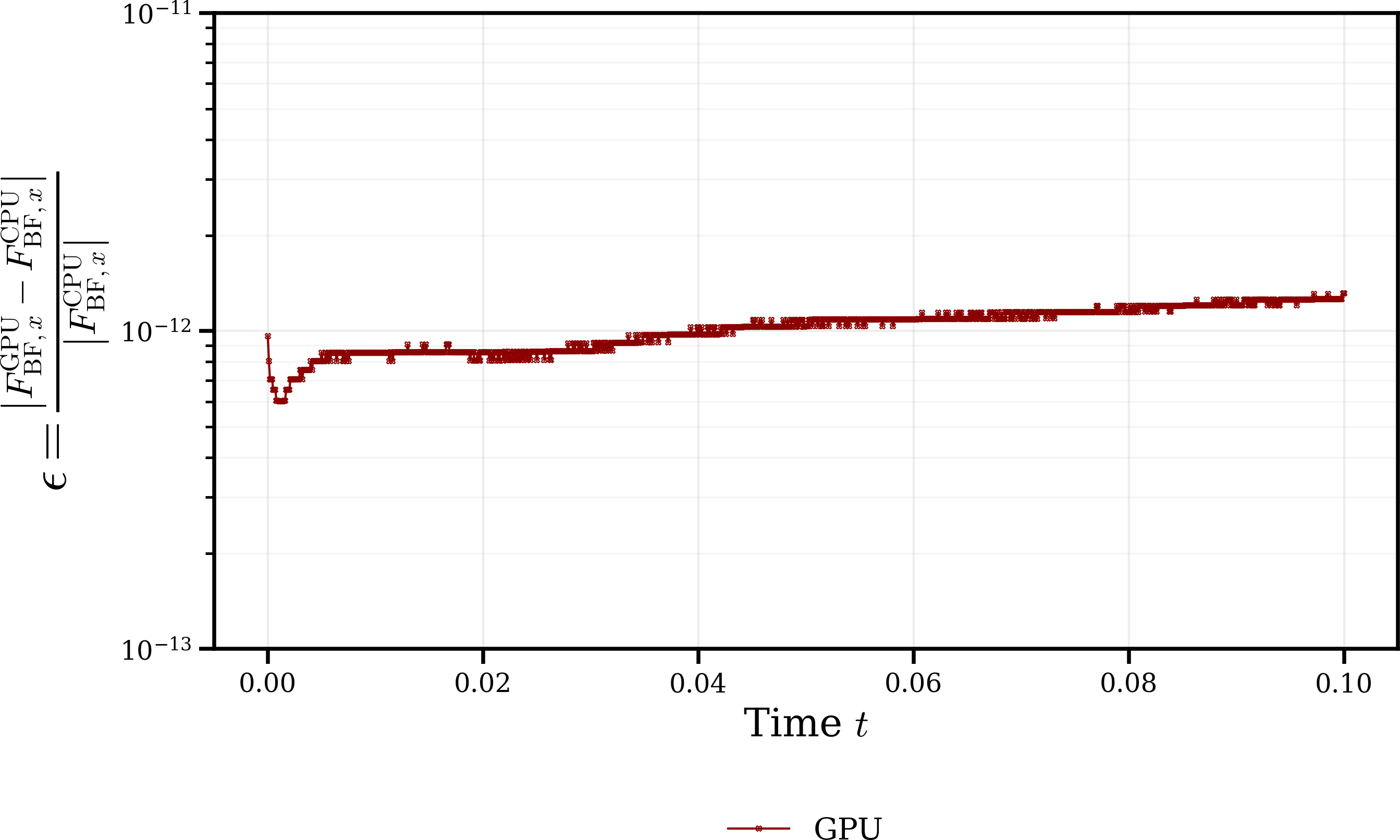}
        \label{fig:wm_relative}
    \end{subfigure}
    \caption{Top: Comparison of instantaneous, streamwise body force $F_{BF,x}$. Bottom: Normalized relative error $\epsilon$ of the GPU implementation with respect to the CPU-based reference.}
    \label{fig:wm_flowfield}
\end{figure}

\Cref{fig:wm_flowfield} shows the initial transient evolution of the instantaneous, streamwise body force $F_{BF,x}$ and the normalized relative deviation $\epsilon$ between the CPU and GPU results of the compressible version of Spalding's wall model. The subscript $(\cdot)_x$ denotes the streamwise ($x$) component of the corresponding variable. The results demonstrate that the relative deviations in the observed streamwise body force remain within machine floating-point precision. This exceptional agreement rigorously validates the correctness and numerical fidelity of the proposed GPU-acceleration strategy.

\section{WRLES of a Transient Compressor Cascade}
\label{sec:fsi}
The interaction between shock waves and turbulent boundary layers is among the most complex flow phenomena encountered in aerodynamics. One application where these interactions are prevalent is the transonic flow through an axial rotor~\cite{meng_large_2024}. As the passage shock intersects the blade suction surface, it imposes a sharp adverse pressure gradient on the boundary layer, promoting transition, thickening, and in severe cases, full flow separation. The effects extend beyond localized aerodynamic losses, with strong shock-turbulent boundary layer interactions being the primary driver of unsteady shock motion that governs aeroelastic response and energy transfer, rendering their accurate characterization fundamental to both compressor performance prediction and efficient blade design.

The NASA Rotor 67 has long served as a canonical benchmark for transonic compressor research, with well-documented experimental data spanning a range of operating conditions \cite{wood_shock_1986, pierzga_investigation_1985, strazisar_investigation_1985, strazisar_laser_1989}. A notable trend shown in literature is the progressive movement of the shock in the passage. The movement is driven by increasing back pressure from the choked flow condition towards near stall condition, pushing the passage shock forwards towards the leading edge. This progression intensifies the effects of the interaction between the shock and the boundary layer on the suction side of the blade.

These shock-boundary layer interactions are examined through a series of 2.5D numerical simulations of the Rotor 67 blade section at $70\%$ span measured from the hub. Simulations were conducted at varying back pressure conditions to track the forward migration of the passage shock as conditions approach stall. In the 2.5D framework, the back pressure required to drive strong STBLI differs from that observed in three-dimensional experiments and a key objective of this work is to establish the revised pressure conditions under which these critical flow features emerge.

All simulations are performed with \galaexi on the Hunter supercomputer. The computational grid comprises $3.6 \times 10^7$ DOFs
at a polynomial degree of $N = 3$. The rotor blade section is treated as an adiabatic no-slip wall, with periodic boundary
conditions enforced at the spanwise extents. The mesh is refined in the near-wall region to ensure wall-resolved LES, achieving an averaged $y^+ < 0.8$ across the blade surface. Inflow is prescribed via a supersonic Dirichlet boundary condition, and the outflow is handled using a subsonic pressure outflow condition. Temporal discretization is achieved through a fourth-order low-storage
Runge--Kutta scheme, with the time step governed by a CFL number of 0.9. Spatial discretization employs a split DGSE formulation with
an entropy conserving flux function~\cite{Chandrashekar2013}. Numerical fluxes at the faces are computed using the Roe Riemann
solver with entropy correction according to Harten et al.~\cite{Harten1983}. The working fluid is modeled as an ideal gas, with dynamic viscosity evaluated using Sutherland's law.
The simulation conditions, summarized in \cref{tab:RotorFlowConditions}, share a relative inflow Mach number of $\mathrm{Ma}_{rel}=1.197$ at a Reynolds number of $\mathrm{Re}=2.6 \times 10^6$ based on the chord length $c=\SI{0.093}{\meter}$. The inputs are based on the three-dimensional experimental values from \cite{strazisar_laser_1989}, with Case 1 being the inputs required for peak efficiency and Case 2 being the inputs for near stall. Freestream inflow conditions are set to the sea level conditions from the International Standard Atmosphere (ISA)~\cite{iso1975isa}.

\begin{table}[h]
\centering
\caption{Simulation inflow and outflow conditions, where $\alpha_{rel}$ is defined as the inflow relative flow angle.}
\label{tab:RotorFlowConditions}
\begin{tabular}{@{}p{1cm}p{1.5cm}p{1.5cm}p{1.5cm}@{}}
\toprule
Case & $\mathrm{Ma}_{rel}$ & $\alpha_{rel}$ & $p_2/p_{ref}$ \\ \midrule
1    & 1.197           & 63.2                  & 1.1862  \\
2    & 1.197           & 64.6                  & 1.3242  \\ \bottomrule
\end{tabular}
\end{table}

\Cref{tab:RotorPratios} presents a comparison of the LES results against experimental data at measurement planes positioned one chord length upstream and downstream of the blade, showing generally good agreement in the static pressure ratios across both conditions, although it is noted that Case 1 does have a noticeable difference at the upstream measurement plane.

\begin{table}[h]
\centering
\caption{Pressure ratios at measurement planes upstream and downstream of the rotor.}
\label{tab:RotorPratios}
\begin{tabular}{@{}lcc@{}}
\toprule
Case              & $p_1/p_{ref}$ & $p_2/p_{ref}$ \\ \midrule
LES Case 1        & 0.91          & 1.182         \\
Experiment Case 1 & 0.79          & 1.186          \\
LES Case 2        & 0.84          & 1.31          \\
Experiment Case 2 & 0.83          & 1.32          \\ \bottomrule
\end{tabular}
\end{table}

\begin{figure}[h!]
    \centering
    \includegraphics[width=0.8\textwidth]{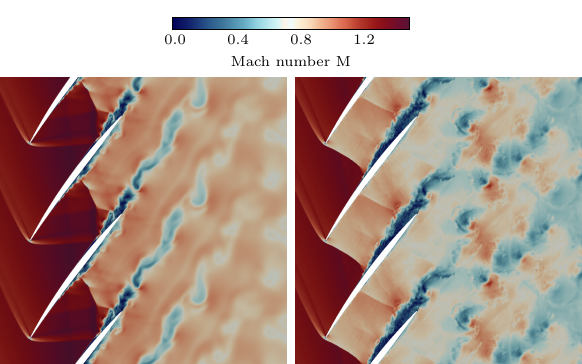}
    \caption{Shock wave formation for a static pressure ratio (Case 2; left) and increased back pressure (right). The instantaneous flow field coloured by Mach number at $t/CTU=50$ is shown.}
    \label{fig:RotorResults}
\end{figure}

At both the peak efficiency and near stall conditions, the passage shock remains in the aft portion of the blade passage,
intersecting the suction surface close to the trailing edge, as shown in the left image of \cref{fig:RotorResults}. This is
inconsistent with the trend observed in the three-dimensional experiments \cite{wood_shock_1986} and the three-dimensional
simulations by \cite{fior_cfd_2018}, where increasing back pressure drives the shock progressively forward into the passage. This
result reinforces the notion that the pressure ratios derived from three-dimensional experimental data do not produce an
equivalent shock structure when applied to a 2.5D blade section simulation, a consequence of the missing three-dimensional flow
effects. For this reason, the outlet pressure was increased beyond the experimental near-stall value until the pressure ratio at the
downstream station was $p_2/p_{ref} = 1.53$. This new back pressure was found to be sufficient to drive the shock forwards and establish a
condition which better resembles the three-dimensional shock structures. Results with the larger back pressure are shown in the right image of \cref{fig:RotorResults}.

The results of the study demonstrate that the 2.5D modelling framework requires a higher back pressure than the experimentally-defined near-stall value. Prescribing a higher back pressure allows the reproduction within the current methodology of the equivalent shock structures observed in three-dimensional experiments. This highlights the significant influence of three-dimensional flow effects, such as tip clearance, spanwise redistribution and passage blockage, in promoting earlier shock migration in the full rotor. In both flow conditions, the shock induces a rapid transition to a turbulent boundary layer, confirming that despite the elevated pressure ratio inherent to the 2.5D approach, the configuration remains well-suited for investigating STBLI and, by extension, the resulting aeroelastic effects. These findings will serve as a basis for future work in which the effects of the unsteady shock motion and the resulting aeroelastic effects will be investigated.

\section{Summary}
\label{sec:summary}
The increased usage of GPU architectures in the exascale computing era allows the computation of ever more complex flow problems. The present work showcases recent advancements in the GPU-accelerated CFD solver \galaexi and its application to such problems. It was shown that \galaexi provides excellent strong scaling up to \num{256} AMD MI300A APUs on the Hunter supercomputer. Subsequent weak scaling runs, utilizing up to \num{512} APUs, demonstrated a parallel efficiency exceeding 90\%. The effect of the three available compute partitioning modes on the AMD MI300A APUs on the performance of \galaexi was also studied. It was demonstrated that CPX mode, where a single APU is treated as six logical GPUs, averages a 10\% reduction in PID across a range of computational loads. Up to a 3x speedup in time-to-solution was also observed when using CPX mode when compared to the default SPX mode.

\galaexi was also extended to a more diverse set of applications with the implementation of features necessary to support WMLES. The GPU-acceleration strategy for this feature is outlined for the case of an equilibrium-based, algebraic wall stress model. Special considerations were given to the interface communication and the boundary condition implementation. A plane turbulent channel flow was chosen as a validation test case. The resulting numerical solutions from the GPU-accelerated solver match the baseline CPU implementation to near machine precision.
Finally, \galaexi was applied to more complex flow phenomena via a WRLES of a 2.5D configuration of the transonic NASA Rotor 67 geometry. The study of different back pressures reveals that the 2.5D simulation requires a greater pressure ratio when compared to experiment to ensure that the three-dimensional shock structures are replicated.

\begin{acknowledgement}
This work was funded by the European Union and has received funding from the European High Performance Computing Joint Undertaking (JU)
and Sweden, Germany, Spain, Greece, and Denmark under grant agreement No 101093393.
This work is partly funded by the Deutsche Forschungsgemeinschaft (DFG, German Research Foundation) with - EXC2075 -- 390740016 under Germany's Excellence Strategy and in the framework of the research unit FOR 2895 and FOR 2687.
The research presented in this paper was funded in parts by the state of Baden-Württemberg under the project Aerospace 2050 MWK32-7531-49/13/7 "FLUTTER"/"QUASAR".
The authors also gratefully acknowledge the Federal Ministry for Economic Affairs and Energy (BMWE) for funding this work in the framework of the research project SaFuMa (FKZ: 20E2409)
We acknowledge the support by the Stuttgart Center for Simulation Science (SimTech).
\end{acknowledgement}

\bibliography{References}
\bibliographystyle{abbrv}

\end{document}